\documentclass[3p,times,twocolumn]{elsarticle}

\usepackage{ecrc}

\volume{00}

\firstpage{1}

\journalname{Nuclear and Particle Physics Proceedings}

\runauth{}

\jid{nppp}

\jnltitlelogo{Nuclear and Particle Physics Proceedings}

\usepackage{amssymb}

\usepackage[figuresright]{rotating}
\usepackage{graphicx}

\begin{document}

\begin{frontmatter}



\dochead{}

\title{SCET for jet physics in the vacuum and the medium}


\author{Ivan Vitev}

\address{Theoretical Division, Los Alamos National Laboratory, Mail Stop B283, Los Alamos, NM 87545, USA}

\begin{abstract}

In this plenary talk I discuss soft-collinear effective theory (SCET) as a framework for precision QCD phenomenology. 
Emphasis is placed on jet and heavy flavor observables accessible at current and future  collider facilities. 
One of the principal challenges that calculations of hard probes in heavy ion reactions face is the ambiguity associated with the implementation of  medium-induced radiative effects.  I demonstrate how extension of SCET to describe  parton propagation in
QCD matter has helped quantify and reduce the theoretical uncertainty in jet quenching calculations. 
\end{abstract}


\end{frontmatter}



\section{Introduction}
\label{intro}

The purpose of these proceedings is to highlight  selected recent results obtained in the framework of soft-collinear 
effective theory (SCET)~\cite{Bauer:2000yr,Beneke:2002ph} 
and its extension to describe jet propagation in a background QCD medium via Glauber gluon 
exchange~\cite{Idilbi:2008vm,Ovanesyan:2011xy}. Emphasis is placed on observables that
illustrate the gains in precision from higher-order calculations and resummation. Observables of direct relevance to  experiments at  
current and future high-energy nuclear physics facilities such as
 the Relativistic Heavy Ion Collider (RHIC), the Large Hadron Collider (LHC) and an Electron Ion Collider (EIC) are given 
 priority.  Results for jets and heavy flavor, strictly within traditional pQCD, are covered 
 elsewhere and summarized in~\cite{Jeon}. Experimental results are collected in~\cite{Caines}.

\section{SCET for jet physics in the vacuum}
\label{sec-med}

Soft-collinear effective theory  has emerged as a new tool to address hard  large $Q^2$ 
processes in lepton-lepton,  lepton-hadron, and hadron-hadron collisions.  
Together with QCD factorization, which has been proven in this framework for a number of
processes, it is  especially useful in improving the precision of multi-scale 
calculations through the resummation of large Sudakov-type logarithms. While initially a large
body of work was dedicated to $e^+e^-$ annihilation, recently there has been more focused effort 
toward processes of  interest to LHC phenomenology and a future EIC. 

\begin{figure}
 \includegraphics[width=8cm,clip]{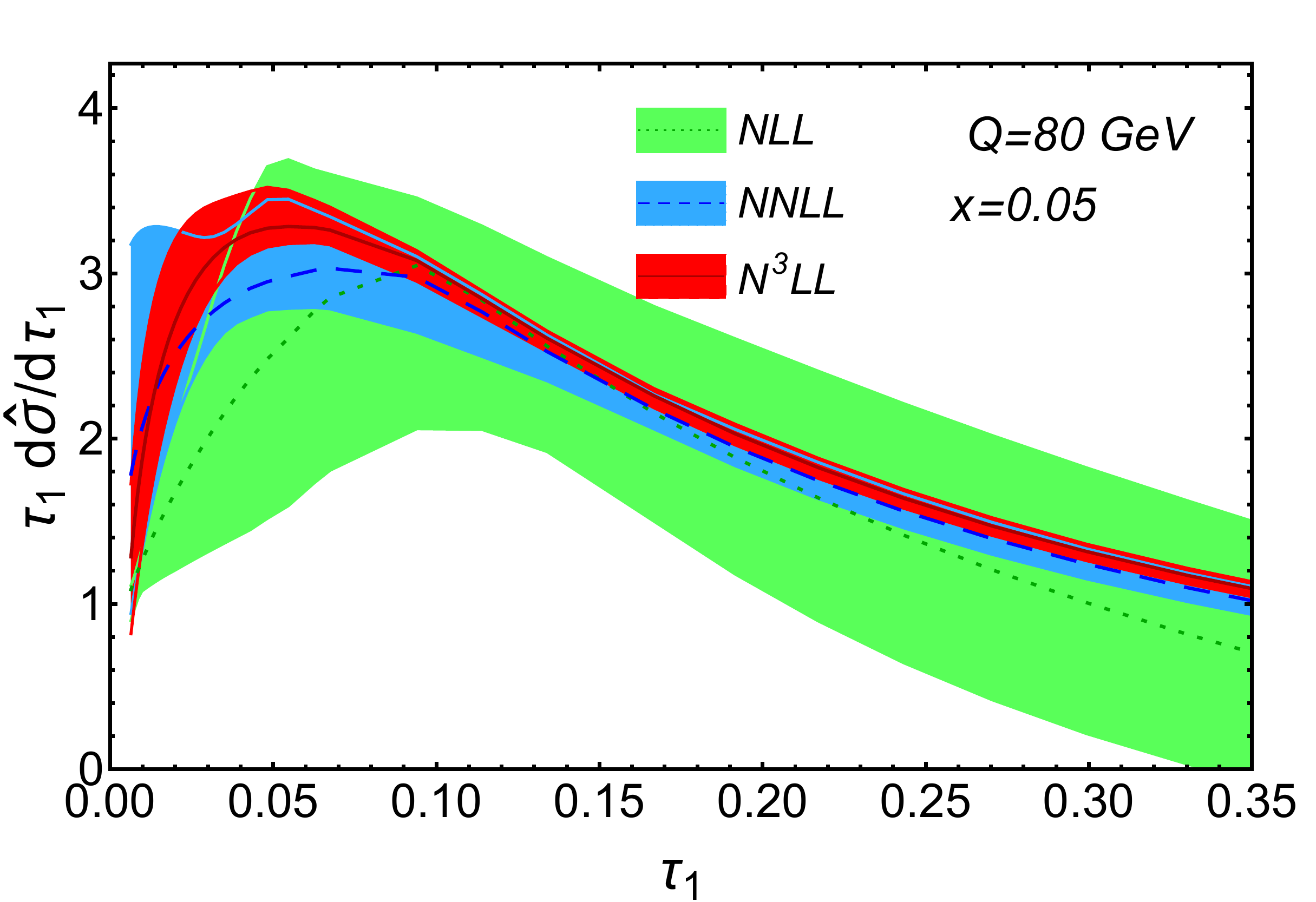} 
\vspace*{-0.5cm}       
\caption{Example of perturbative convergence of the 1-jettiness for NLL, NNLL, N$^3$LL resummation at $x=0.05$ 
and $Q=80$~GeV~\cite{Kang:2015swk}. }
\label{fig-tau}       
\end{figure}

As a first example we consider one inclusive jet production in deep inelastic scattering (DIS).  
The discussed observable is called 1-jettiness in DIS is defined by one jet and one
beam axis
\begin{equation}
\tau_1 \equiv \frac{2}{Q^2}\sum_{i\in X}\min\{ q_B \cdot p_i, q_J\cdot p_i\}\,.
\end{equation}
Here $q_B,q_J$ are lightlike four-vectors along the beam and jet directions. 
In terms of collimated parton shower structures, this is similar to 2-jettines in $e^+e^-$  (two 
jets in the final state) and 0-jettiness in $pp$ (Drell-Yan).
This event shape distribution has been calculated over the past 15 years with increasing  
theoretical precision from next-to-leading logarithmic (NLL) accuracy to next-to-next-to-next-to-leading
(N$^3$LL) logarithmic accuracy~\cite{Antonelli:1999kx,Kang:2014qba,Kang:2013lga,Kang:2015swk}.  
An example of these improvements can be seen in Fig.~\ref{fig-tau}~\footnote{The figure is reproduced with 
the authors' permission.}. The uncertainty band is reduced from a factor of few to just a  few percent. 
The kinematics is chosen to be representative of HERA measurements. These improvements can help test the universality
of non-perturbative effects and extract the strong coupling constant at the $Z^0$ pole, $\alpha_s(m_Z)$. 
The development of this technology is also useful in broadening the scope of the future EIC physics
program.    

\begin{figure}
 \includegraphics[width=8.5cm,clip]{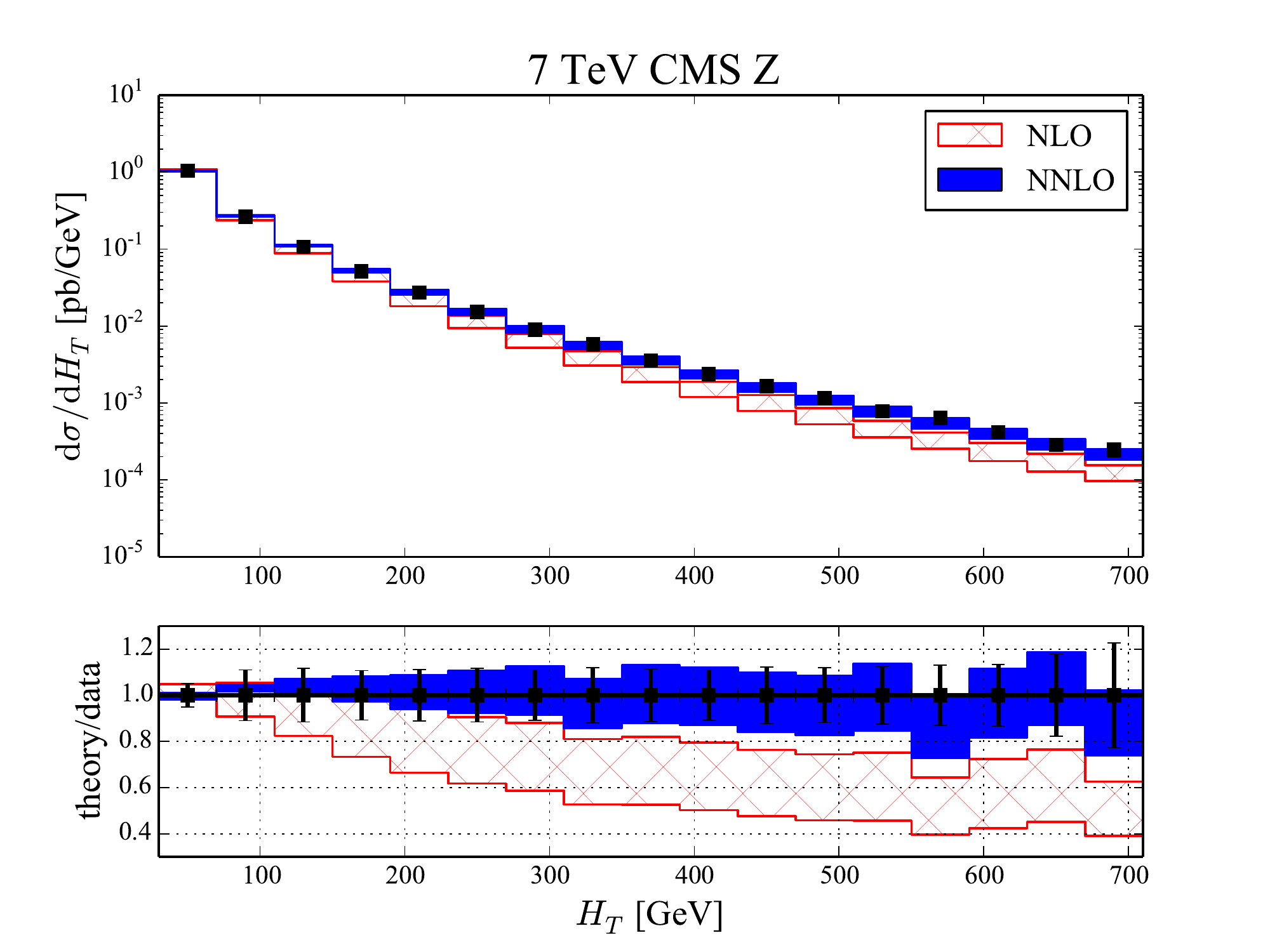} 
\vspace*{-0.5cm}       
\caption{  The scalar sum of jet transverse momenta distribution in Z+jet processes measured by CMS.  The
 ratio of the NLO and NNLO predictions to the measured data is shown~\cite{Boughezal:2016yfp}.}
\label{fig-Z0jet}       
\end{figure}

In the past few years there has been a proliferation of NNLO calculations for the LHC 
($H$+jet, $W^\pm/Z^0$+jet, etc). While NLO $V$+N jet calculations, that can also be matched to parton showers, generally 
work well, there are notable exceptions.  One such example is the scalar momentum sum $p_T$ distributions of associated
jets. One of the main challenges in such calculations is the treatment of infrared (IR) singularities. Generally, two approaches 
are commonly adopted, local and non-local subtraction schemes. SCET, and the N-jettiness variable has found a
novel application in a non-local subtraction scheme~\cite{Boughezal:2015aha,Boughezal:2016yfp}. At NNLO the cross section can have up to two additional partons in the
final state and can be expressed schematically as follows 
 \begin{eqnarray}
\label{eq:partition}
\hspace*{-0.9cm}\sigma_{NNLO}  \!\!\! \!\! &=&\!\!\! \! \!  \int {\rm d}\Phi_N \, |{\cal M}_{N}|^2 +\int {\rm d}\Phi_{N+1} \, |{\cal M}_{N+1}|^2 \, \theta_N^{<} 
\nonumber \\
 &&  \!\!\!  \!\! \!  \! +\int {\rm d}\Phi_{N+2} \, |{\cal M}_{N+2}|^2 \, \theta_N^{<}+\int {\rm d}\Phi_{N+1} \, |{\cal M}_{N+1}|^2 \, \theta_N^{>}
\nonumber  \\
&&  \!\!\!  \!\! \!  \! +\int {\rm d}\Phi_{N+2} \, |{\cal M}_{N+2}|^2 \, \theta_N^{>}\, . 
\end{eqnarray}
Here, $\theta_N^{>}$ and $\theta_N^{<}$ represent a cut for a small value of the N-subjettiness variable $\tau_N$. Below $\tau_N$
one uses the factorization theorems of SCET to evaluate the cross section.  If the cross section is expanded to the appropriate fixed order it will reproduce the NNLO result. Above  $\tau_N$ the fixed order calculation works well, in particular one needs the results with
N+1 and N+2 jets. An example of the calculation of the scalar sum of transverse momenta of jets associated with a $Z^0$ 
boson is shown in Fig.~\ref{fig-Z0jet}~\footnote{The figure is reproduced with 
the authors' permission.}. The theoretical uncertainties are reduced and there is much better  agreement between theory 
and experiment.

A noteworthy development in the past year was the development of SCET resummation for semi-inclusive
jet observables. For jet production, logarithms of the jet radius parameter arise. When the jet radius $R$ is small, such  logarithms can become large  and require resummation. It was recently shown that when the out-of-cone radiation is not power
suppressed, ${\cal O}(\Lambda/E_J)$,  these terms are of the form $\left(\alpha_s \ln R\right)^n$~\cite{Idilbi:2016hoa,Kang:2016mcy,Kang:2016ehg,Dai:2016hzf}. The new semi-inclusive jet functions follow DGLAP-type evolution 
equations~\cite{Altarelli:1977zs}.  For more details see the contribution by Ringer~\cite{Ringer}.

\section{SCET for jet physics in a QCD medium}
\label{sec-med}

An effective field theory (EFT) for hard processes in heavy ion collisions can be constructed by coupling the jets to  the QCD medium by off-shell $t$-channel Glauber gluon exchanges with momentum scaling  
$q \sim (\lambda^2, \lambda^2, {\bf \lambda })$, where $\lambda$ is a small parameter. Building upon the soft-collinear effective theory  of jet  production~\cite{Bauer:2000yr}, the collinear  quark-Glauber and collinear gluon-Glauber sectors of the extended theory   SCET${_{\rm G}}$ were  derived in \cite{Idilbi:2008vm} and \cite{Ovanesyan:2011xy}. In this background field approach, the properties of the QCD medium that determine the jet-medium interactions  enter the potential that sources the Glauber gluons and first applications discussed the transverse momentum broadening 
of jets~\cite{Idilbi:2008vm,Ovanesyan:2011xy,DEramo:2010wup}. A critical step that enables calculations of parton shower modification
in strongly-interacting matter and applications to jet quenching  phenomenology beyond the traditional energy loss approach is  the derivation of the  
all four splitting functions  $q\rightarrow qg$,  $g\rightarrow gg, g\rightarrow q\bar{q}$ and $q\rightarrow gq$  to first order in opacity~\cite{Ovanesyan:2011kn}. The effect of correction that arise from the finite parton scattering kinematics, branching kinematics, and recoil of the constituents of the QCD medium has been studied.

On of the principal challenges that faces the  theory of hard probes  in heavy ion reactions is the unified description of  vacuum and in-medium
parton showers. The traditional energy loss approach relies on more than two decades of theoretical  and phenomenological  studies. 
In contrast, the 
implementation  of  the medium-induced parton branching requites new strategies and techniques,  some  of which have been  developed in the context
of high energy physics. The benefits of such advances, however, cannot be understated.  Significant  improvements in the precision of heavy
ion phenomenology from resummation and higher order corrections~\cite{Almeida:2014uva}  are expected. As a first example, the corrections  
to the  DGLAP evolution equations~\cite{Altarelli:1977zs} in the QGP were considered, with SCET$_{\rm G}$ splitting kernels as 
input~\cite{Kang:2014xsa}. This allowed us to quantify the uncertainties  due to the implementation of the in-medium modification to inclusive 
hadron production. Predictions for light hadron  suppression~\cite{Chien:2015vja}  are in excellent agreement with preliminary 
CMS results~\cite{Baty}.

\begin{figure}
 \includegraphics[width=8.cm,clip]{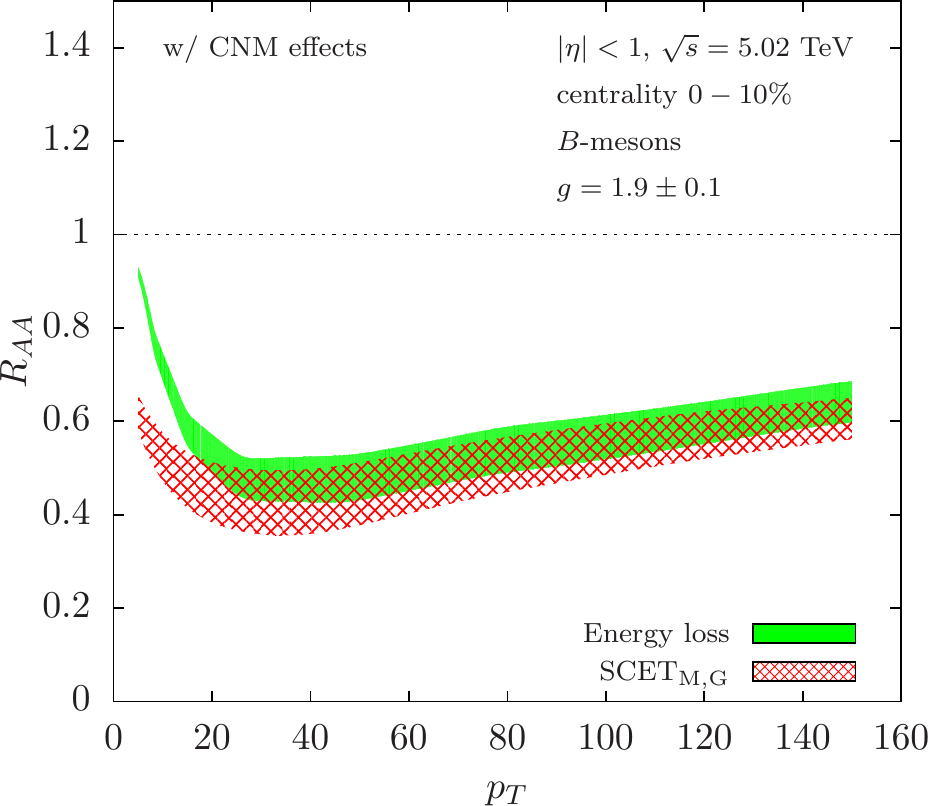} 
\vspace*{-0.3cm}       
\caption{ The nuclear modification factor $R_{AA}$ for $B^+$ meson (right) production as a function of the transverse momentum $p_T$. Result obtained within the traditional approach to energy loss are compared to SCET$_{\rm M,G}$-based calculations~\cite{Kang:2016ofv}. }
\label{fig-comparison}       
\end{figure}

To extend the in-medium EFT approach to the heavy flavor sector,  one  needs to to couple the charm and beauty quarks to the
QCD medium. This was recently done in~\cite{Kang:2016ofv} and reported at this conference by Ringer~\cite{Ringer}.
The SCET$_{\rm M}$ Lagrangian in the vacuum with quark masses was obtained  in~\cite{Leibovich:2003jd}. Introduction of heavy 
quark masses requires specific power counting,  $m/p^+ \sim \lambda$  of the order of the small power counting parameter in SCET. This is also 
consistent with the power counting for the dominant transverse momentum component of the Glauber gluon exchange. 
 Hence, to lowest order the new effective  theory of heavy quark propagation in matter. SCET$_{\rm M,G}$ = 
 SCET$_{\rm M}\otimes$SCET$_{\rm G}$. The three splitting processes where the heavy quark mass plays a role, $Q\to Qg$, $Q\to gQ$ 
 and $g\to Q\bar Q$, have been computed analytically to first order in opacity and evaluated numerically.
 Incorporating their contribution in a framework consistent with next-to-leading (NLO) calculation can be schematically expressed as
\begin{equation}
\label{eq:AA}
d\sigma^H_{{\rm PbPb}} = d\sigma^{H, {\rm NLO}}_{pp} + d\sigma^{H, {\rm med}}_{{\rm PbPb}},
\end{equation}
where $d\sigma^{H, {\rm NLO}}_{pp}$ is the NLO cross section in the vacuum, and $d\sigma^{H, {\rm med}}_{{\rm PbPb}}  
= \hat \sigma^{(0)}_i \otimes D_i^{H,{\rm med} }$ is the one-loop medium correction. Comparison of B-meson $R_{AA}$ using 
the NLO  framework of~\cite{Jager:2002xm} and the fragmentation functions of~\cite{Kniehl:2008zza} and the suppression 
obtained using the  traditional energy loss approach is shown in Fig.~\ref{fig-comparison}. At high $p_T$ the two approaches agree
within the theoretical uncertainties. At lower $p_T$ the significant gluon fragmentation contribution to open heavy flavor leads to 
values of $R_{AA}$ that may be smaller by as much as 50\%.

\begin{figure}
 \includegraphics[width=8.cm,clip]{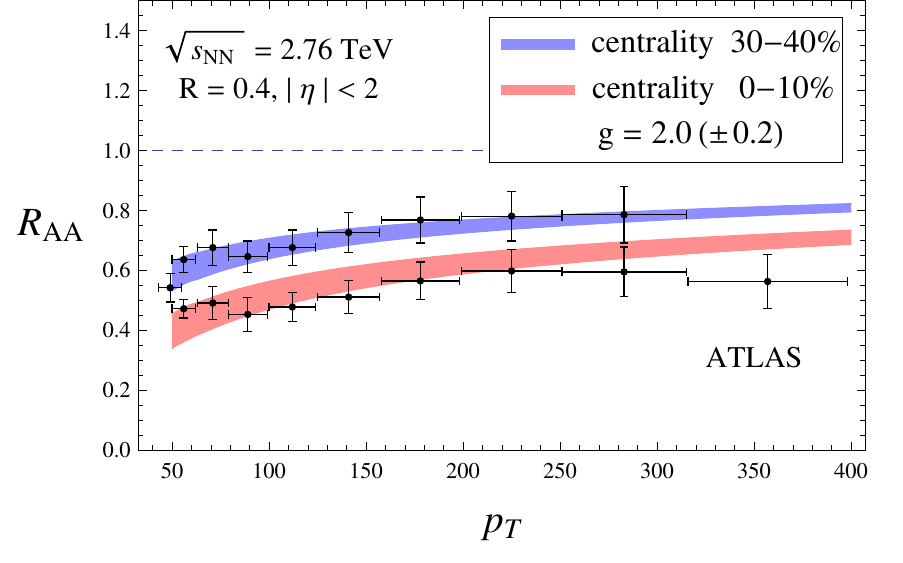} 
\vspace*{-0.3cm}       
\caption{Calculations of the nuclear modification factor $R_{AA}$ of inclusive jets as a function of the jet transverse momentum are compared to experimental data in central and mid-peripheral Pb+Pb collisions at $\sqrt{s}=2.76$~TeV at the LHC. Bands correspond to the theoretical uncertainty estimated by varying the coupling between the jet and the medium ($g=2.0 \pm 0.2$)~\cite{Chien:2015hda}. }
\label{fig-jetRAA}       
\end{figure}

The transverse and longitudinal structure differences between the vacuum and in-medium parton showers can be
clarified in considerable detail through studies  of jets and jet substructure in proton and heavy ion collisions. Since
jets are defined through a reconstruction  algorithm with jet radius parameter $R$, the concept of energy loss can be
generalized due to out-of-cone radiation~\cite{Chien:2015hda}.  More specifically, the medium-induced energy loss of a quark or gluon initiated 
jets is
\begin{eqnarray}
    \epsilon_i
   & =& \frac{2}{\omega}\Big[\int_0^{\frac{1}{2}} dx k^0
    +\int_{\frac{1}{2}}^1 dx (p^0-k^0)
    \Big] \nonumber \\ 
&&    \int_{\omega x(1-x)\tan\frac{R}{2}}^{\omega x(1-x)\tan\frac{R_0}{2}}dk_\perp   \frac{1}{2}  \sum_i {\cal P}^{med}_{i\rightarrow jk}(x,k_\perp)   \;.
\end{eqnarray}
Here, $R$ is the angular parameter used in the jet reconstruction, and $R_0$ is of ${\cal O}(1)$ in QCD, which sets the region of the 
use of collinear parton splitting functions. ${\cal P}^{med}_{i\rightarrow jk}(x,k_\perp)$ are the medium-induced Altarelli-Parisi splitting kernels.  With an emphasis on a the consistent theoretical descriptions of hadron and jet observables in heavy ion collisions,  the 
results presented in~\cite{Chien:2015hda}  include cold nuclear matter (CNM) effects and splitting functions used to evaluate light
particle quenching. In Fig.~\ref{fig-jetRAA} the suppression of jet production in central and mid-peripheral lead-lead collisions is 
compared to ATLAS experimental data~\cite{Aad:2014bxa}.

\begin{figure}
 \includegraphics[width=8.cm,clip]{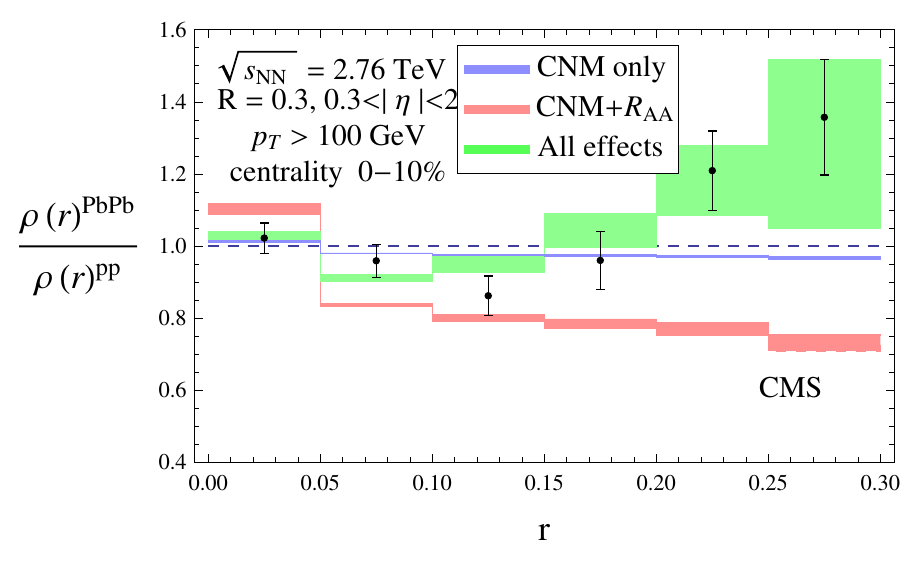} 
\vspace*{-0.3cm}       
\caption{Theoretical calculations for the modification of differential jet shapes of inclusive jets in central Pb+Pb  collisions at the LHC.  The jet transverse momentum $p_T > 100$~GeV and pseudo-rapidity $0.3<|\eta|<2.0$. The coupling between the jet and the medium is fixed at $g=2$. }
\label{fig-shape}       
\end{figure}

The jet shape, a classic substructure observable, can give complimentary information on the in-medium parton shower.
The integral and differential jet shapes are defined as follows
\begin{equation}
    \Psi_J(r)=\frac{\sum_{i,~d_{i\hat n}<r} E^i_T}{\sum_{i,~d_{i\hat n}<R} E^i_T}\;, \quad  \rho(r)=\frac{d}{dr}\Psi(r) \;.
\end{equation}
It was found that the non-trivial behavior of the jet shape modification is caused by both the different quark and gluon jet cross section suppressions and the jet-by-jet broadening. The cross section of gluon-initiated jets is more suppressed, which enhances the fraction of quark-initiated jets having a narrower energy profile. This causes the attenuation of the jet shape in the mid $r$ region. On the other hand, the broadening of jets results in the enhancement of the jet shape near the periphery of the jet. The calculation provides for the first time a quantitative description of the jet shape modification in Pb+Pb collisions at the LHC~\cite{Chatrchyan:2013kwa}, shown in Fig.~\ref{fig-shape}.  

\begin{figure}
\hspace*{-.3cm} \includegraphics[width=8.cm,clip]{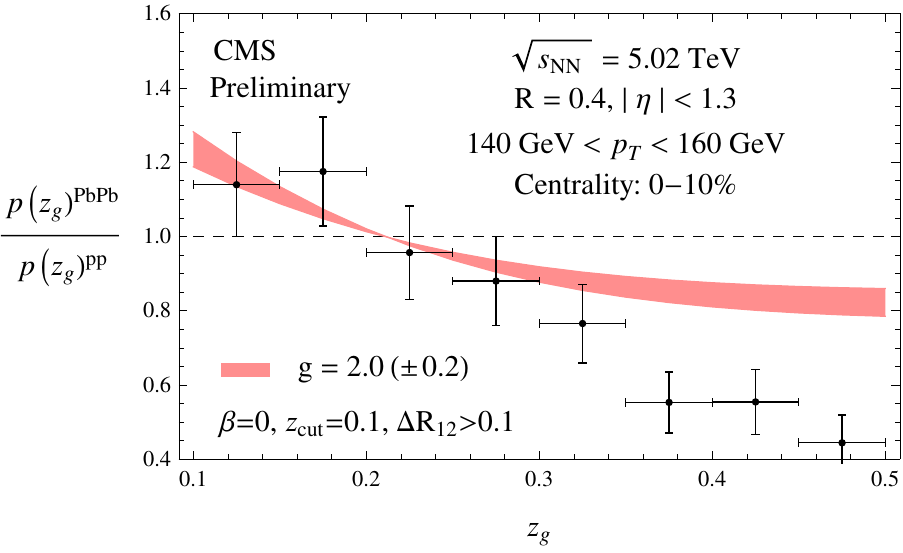} 
\vspace*{-0.5cm}       
\caption{Calculations and preliminary CMS data for the ratio of momentum sharing distributions of inclusive anti-$k_T$ $R=0.4$ jets in  central Pb+Pb and p+p collisions at $\sqrt {s_{\rm NN}} = 5.02$ TeV. The band corresponds to the theoretical uncertainty estimated by varying the coupling between the jet and the medium ($g=2.0\pm0.2$)~\cite{Chien:2016led}.} \label{fig-zg1}       
\end{figure}

Deeper inspection of jets has recently become possible~\cite{Larkoski:2014wba}. In the context of heavy ion collisions, the 
groomed soft-dropped momentum sharing and angular separation distributions between the leading subjets 
inside a reconstructed jet are particularly interesting~\cite{Chien:2016led}.  These observables are directly sensitive to the hardest branching in the process of jet formation and are an ideal tool to study the early stage of the in-medium parton shower evolution. To construct these variables
one goes through the branching history of a shower, eliminating the soft branch at each step until 
\begin{equation}
    z_{cut} < \frac{\min(p_{T_1},p_{T_2})}{p_{T_1}+p_{T_2}} \equiv z_g\;. 
    \label{SD}
\end{equation}
A further minimum angular separation restriction between the two branches   $\Delta < \Delta R_{12} \equiv r_g$ is imposed, where 
$\Delta R_{12}$ isdefined as the groomed jet radius $r_g$.  Examination of the  modification of the momentum sharing 
distribution $p(z_g)$ and the groomed radius distribution  $p(r_g)$ distribution can shed light on the parton shower modification 
 in heavy ion collisions.  Specifically, one can select the jet transverse momentum and the angle between the two leading subjets, ensure large splitting virtuality and, consequently, a branching which happens shortly after the hard scattering in the QGP. The  branching time, estimated as follows 
 \begin{equation}
\tau_{\rm br} [{\rm fm}] = \frac{0.197\; {\rm GeV~fm} }{z_g(1-z_g) \,  \omega [{\rm GeV}]  \, \tan^2( r_g / 2)} \,  ,
\end{equation}
suggests that for typical jets with $\omega = 2 p_T  = 400$~GeV, $r_g = 0.1$ and $z_g = 0.1$, the branching 
time $\tau_{\rm br}  < 2$~fm. 
This selects early splitting process inside the QGP of size $\sim 10$~fm created in
Pb+Pb collisions at the LHC.  The modification of the momentum sharing and angular separation distributions in lead-lead relative to proton-proton collisions is evaluated using the leading-order medium-induced splitting functions obtained in the framework of soft-collinear effective theory with Glauber gluon interactions~\cite{Ovanesyan:2011kn}. Qualitative and in most cases quantitative agreement between theory and preliminary CMS measurements~\cite{Caines} is observed.  The medium enhances the asymmetric 
branching and one example in central heavy ion reactions at the LHC is
shown in Fig.~\ref{fig-zg1} for the jet momentum range $140~{\rm GeV}< p_T < 160~{\rm GeV}$.  The variance at $z_g \sim 0.5$ between data and theory disappears at higher $p_T$ and less central collisions. The STAR collaboration at RHIC did not see strong groomed jet modification at lower transverse momenta.

\begin{figure}
 \includegraphics[width=8.5cm,clip]{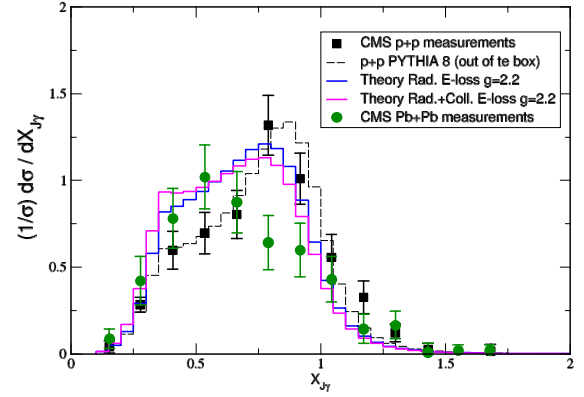} 
\vspace*{-1cm}       
\caption{  The momentum imbalance distribution $X_{J\gamma}$ in proton and heavy ion collisions at
center-of-mass energy  2.76~TeV is compared to preliminary CMS data.  Theoretical calculations 
of the momentum imbalance shift from p+p to Pb+Pb are performed in the energy loss 
limit~\cite{KVX2017} for 0-30\% central collisions.}
\label{fig-Zjet}       
\end{figure}

Finally, it important to note that in the soft gluon emission limit only two of the four medium-induced splitting
intensities survive $q\rightarrow qg$,  $g\rightarrow gg$. This allows for a standard energy loss interpretation of jet 
quenching,  i.e. leading partons  lose energy  through non-Abelian bremsstrahlung and flavor changing processes are  
suppressed~\cite{Ovanesyan:2011kn}.  In this limit, using vector boson ($\gamma$, $Z^0$) tagging, more accurate 
constraints can be placed on jet energy loss in comparison to inclusive jet, and even di-jet measurements. 
Convenient variables that encode such information are the  tagged jet momentum asymmetry $A_{JV}$ and momentum
imbalance  $X_{JV}$m defined as 
\begin{equation}
A_{JV} = \frac{p_{TJ}-p_{TV}}{p_{TJ}+p_{TV}}\;,  \qquad X_{JV}= \frac{p_{TJ}}{p_{TV}}\;.
\end{equation} 
Complete characterization of the  quenching of  photon-tagged jets, for example,  requires measurements of the double 
differential 
distribution and is not possible at present due to limited statistics. Their momentum imbalance can be obtained as follows   
 \begin{equation}
\hspace*{-0.4cm} \frac{d\sigma}{dX_{J\gamma}} = \int_{p_{TJ}^{min}}^{p_{TJ}^{max}}d p_{TJ}
\,  \frac{p_{TJ}}{X_{J\gamma}^2}
\frac{d\sigma[X_{J\gamma},p_{T\gamma}(X_{J\gamma},p_{TJ})]}{dp_{T\gamma}dp_{TJ}} \;,
\end{equation}
where the experimental transverse momentum cuts determine the range of $X_{J\gamma}$
and influence its distribution. The normalized momentum imbalance distribution $({1}/{\sigma})
{d\sigma}/{dX_{J\gamma}}$ is given in Fig.~\ref{fig-Zjet}. Theoretical results~\cite{KVX2017} 
use Pythia8  to simulate the p+p baseline.  Radiative and collisional energy losses are
further included in the description of the photon-tagged jet momentum imbalance in
central Pb+Pb collisions at the LHC. Preliminary CMS data is also included and the
downshift in the $X_{J\gamma}$ distributions from  is qualitatively consistent with  theoretical 
expectations with new results from ATLAS and CMS expected to appear soon~\cite{Caines}.   

\section{Conclusions}
This plenary talk highlighted selected recent results obtained in the framework  of soft-collinear effective theory and its 
extension to include Glauber gluon interactions between jets and strongly-interacting matter.  Precision calculations
of jet production in DIS can help constrain the strong coupling constant and parton distribution functions from existing HERA and 
future EIC data. In hadronic  collisions, tagged jet cross section at NNLO set a remarkably accurate baseline for the study the 
effects of the QCD medium in heavy ion collisions. Small jet radius resummation for inclusive jets and jet substructure 
was recently achieved  in SCET.  An important development for ultra-relativistic nuclear collisions, seeded
by  SCET${_{\rm G}}$,  is the unified treatment  of vacuum and medium-induced parton showers. The advancements in theory
include medium-modified evolution, an extension of the effective theory of jet propagation in QCD medium to heavy quarks, and a 
consistent framework to calculate hard processes in heavy-ion collisions at NLO.  
Significant progress has been achieved in the description of inclusive and tagged jet cross sections and jet substructure  
in heavy ion collisions. \\

{\bf Acknowledgments}
This research was supported by the US Department of Energy, Office of Science under Contract No. DE-AC52-06NA25396 and the DOE Early Career Program.




\nocite{*}
\bibliographystyle{elsarticle-num}
\bibliography{bibliography}

\begin{thebibliography}{100}
\expandafter\ifx\csname url\endcsname\relax
  \def\url#1{\texttt{#1}}\fi
\expandafter\ifx\csname urlprefix\endcsname\relax\def\urlprefix{URL }\fi
\expandafter\ifx\csname href\endcsname\relax
  \def\href#1#2{#2} \def\path#1{#1}\fi

\bibitem{Bauer:2000yr}
C.~W. Bauer, S.~Fleming, D.~Pirjol, I.~W. Stewart, {An Effective field theory
  for collinear and soft gluons: Heavy to light decays}, Phys. Rev. D63 (2001)
  114020.
\newblock \href {http://arxiv.org/abs/hep-ph/0011336}
  {\path{arXiv:hep-ph/0011336}}, \href
  {http://dx.doi.org/10.1103/PhysRevD.63.114020}
  {\path{doi:10.1103/PhysRevD.63.114020}}.

\bibitem{Beneke:2002ph}
M.~Beneke, A.~P. Chapovsky, M.~Diehl, T.~Feldmann, {Soft collinear effective
  theory and heavy to light currents beyond leading power}, Nucl. Phys. B643
  (2002) 431--476.
\newblock \href {http://arxiv.org/abs/hep-ph/0206152}
  {\path{arXiv:hep-ph/0206152}}, \href
  {http://dx.doi.org/10.1016/S0550-3213(02)00687-9}
  {\path{doi:10.1016/S0550-3213(02)00687-9}}.

\bibitem{Idilbi:2008vm}
A.~Idilbi, A.~Majumder, {Extending Soft-Collinear-Effective-Theory to describe
  hard jets in dense QCD media}, Phys. Rev. D80 (2009) 054022.
\newblock \href {http://arxiv.org/abs/0808.1087} {\path{arXiv:0808.1087}},
  \href {http://dx.doi.org/10.1103/PhysRevD.80.054022}
  {\path{doi:10.1103/PhysRevD.80.054022}}.

\bibitem{Ovanesyan:2011xy}
G.~Ovanesyan, I.~Vitev, {An effective theory for jet propagation in dense QCD
  matter: jet broadening and medium-induced bremsstrahlung}, JHEP 06 (2011)
  080.
\newblock \href {http://arxiv.org/abs/1103.1074} {\path{arXiv:1103.1074}},
  \href {http://dx.doi.org/10.1007/JHEP06(2011)080}
  {\path{doi:10.1007/JHEP06(2011)080}}.

\bibitem{Jeon}
S.~Jeon, {Theory summary}, these proceedings.

\bibitem{Caines}
H.~Caines, {Experimental summary}, these proceedings.

\bibitem{Antonelli:1999kx}
V.~Antonelli, M.~Dasgupta, G.~P. Salam, {Resummation of thrust distributions in
  DIS}, JHEP 02 (2000) 001.
\newblock \href {http://arxiv.org/abs/hep-ph/9912488}
  {\path{arXiv:hep-ph/9912488}}, \href
  {http://dx.doi.org/10.1088/1126-6708/2000/02/001}
  {\path{doi:10.1088/1126-6708/2000/02/001}}.

\bibitem{Kang:2014qba}
D.~Kang, C.~Lee, I.~W. Stewart, {Analytic calculation of 1-jettiness in DIS at
  $ \mathcal{O}\left({\alpha}_s\right) $}, JHEP 11 (2014) 132.
\newblock \href {http://arxiv.org/abs/1407.6706} {\path{arXiv:1407.6706}},
  \href {http://dx.doi.org/10.1007/JHEP11(2014)132}
  {\path{doi:10.1007/JHEP11(2014)132}}.

\bibitem{Kang:2013lga}
Z.-B. Kang, X.~Liu, S.~Mantry, {1-jettiness DIS event shape: NNLL+NLO results},
  Phys. Rev. D90~(1) (2014) 014041.
\newblock \href {http://arxiv.org/abs/1312.0301} {\path{arXiv:1312.0301}},
  \href {http://dx.doi.org/10.1103/PhysRevD.90.014041}
  {\path{doi:10.1103/PhysRevD.90.014041}}.

\bibitem{Kang:2015swk}
D.~Kang, C.~Lee, I.~W. Stewart, {DIS Event Shape at N3LL}, PoS DIS2015 (2015)
  142.

\bibitem{Boughezal:2015aha}
R.~Boughezal, C.~Focke, W.~Giele, X.~Liu, F.~Petriello, {Higgs boson production
  in association with a jet at NNLO using jettiness subtraction}, Phys. Lett.
  B748 (2015) 5--8.
\newblock \href {http://arxiv.org/abs/1505.03893} {\path{arXiv:1505.03893}},
  \href {http://dx.doi.org/10.1016/j.physletb.2015.06.055}
  {\path{doi:10.1016/j.physletb.2015.06.055}}.

\bibitem{Boughezal:2016yfp}
R.~Boughezal, X.~Liu, F.~Petriello, {A comparison of NNLO QCD predictions with
  7 TeV ATLAS and CMS data for $V$+jet processes}, Phys. Lett. B760 (2016)
  6--13.
\newblock \href {http://arxiv.org/abs/1602.05612} {\path{arXiv:1602.05612}},
  \href {http://dx.doi.org/10.1016/j.physletb.2016.06.032}
  {\path{doi:10.1016/j.physletb.2016.06.032}}.

\bibitem{Idilbi:2016hoa}
A.~Idilbi, C.~Kim, {Factorization of Jet Mass Distribution with Small
  Radius}\href {http://arxiv.org/abs/1606.05429} {\path{arXiv:1606.05429}}.

\bibitem{Kang:2016mcy}
Z.-B. Kang, F.~Ringer, I.~Vitev, {The semi-inclusive jet function in SCET and
  small radius resummation for inclusive jet production}, JHEP 10 (2016) 125.
\newblock \href {http://arxiv.org/abs/1606.06732} {\path{arXiv:1606.06732}},
  \href {http://dx.doi.org/10.1007/JHEP10(2016)125}
  {\path{doi:10.1007/JHEP10(2016)125}}.

\bibitem{Kang:2016ehg}
Z.-B. Kang, F.~Ringer, I.~Vitev, {Jet substructure using semi-inclusive jet
  functions in SCET}, JHEP 11 (2016) 155.
\newblock \href {http://arxiv.org/abs/1606.07063} {\path{arXiv:1606.07063}},
  \href {http://dx.doi.org/10.1007/JHEP11(2016)155}
  {\path{doi:10.1007/JHEP11(2016)155}}.

\bibitem{Dai:2016hzf}
L.~Dai, C.~Kim, A.~K. Leibovich, {Fragmentation of a Jet with Small Radius},
  Phys. Rev. D94~(11) (2016) 114023.
\newblock \href {http://arxiv.org/abs/1606.07411} {\path{arXiv:1606.07411}},
  \href {http://dx.doi.org/10.1103/PhysRevD.94.114023}
  {\path{doi:10.1103/PhysRevD.94.114023}}.

\bibitem{Altarelli:1977zs}
G.~Altarelli, G.~Parisi, {Asymptotic Freedom in Parton Language}, Nucl. Phys.
  B126 (1977) 298--318.
\newblock \href {http://dx.doi.org/10.1016/0550-3213(77)90384-4}
  {\path{doi:10.1016/0550-3213(77)90384-4}}.

\bibitem{Ringer}
F.~Ringer, {Jet and heavy flavor production in heavy-ion collisions}, these
  proceedings.

\bibitem{DEramo:2010wup}
F.~D'Eramo, H.~Liu, K.~Rajagopal, {Transverse Momentum Broadening and the Jet
  Quenching Parameter, Redux}, Phys. Rev. D84 (2011) 065015.
\newblock \href {http://arxiv.org/abs/1006.1367} {\path{arXiv:1006.1367}},
  \href {http://dx.doi.org/10.1103/PhysRevD.84.065015}
  {\path{doi:10.1103/PhysRevD.84.065015}}.

\bibitem{Ovanesyan:2011kn}
G.~Ovanesyan, I.~Vitev, {Medium-induced parton splitting kernels from Soft
  Collinear Effective Theory with Glauber gluons}, Phys. Lett. B706 (2012)
  371--378.
\newblock \href {http://arxiv.org/abs/1109.5619} {\path{arXiv:1109.5619}},
  \href {http://dx.doi.org/10.1016/j.physletb.2011.11.040}
  {\path{doi:10.1016/j.physletb.2011.11.040}}.

\bibitem{Almeida:2014uva}
L.~G. Almeida, S.~D. Ellis, C.~Lee, G.~Sterman, I.~Sung, J.~R. Walsh,
  {Comparing and counting logs in direct and effective methods of QCD
  resummation}, JHEP 04 (2014) 174.
\newblock \href {http://arxiv.org/abs/1401.4460} {\path{arXiv:1401.4460}},
  \href {http://dx.doi.org/10.1007/JHEP04(2014)174}
  {\path{doi:10.1007/JHEP04(2014)174}}.

\bibitem{Kang:2014xsa}
Z.-B. Kang, R.~Lashof-Regas, G.~Ovanesyan, P.~Saad, I.~Vitev, {Jet quenching
  phenomenology from soft-collinear effective theory with Glauber gluons},
  Phys. Rev. Lett. 114~(9) (2015) 092002.
\newblock \href {http://arxiv.org/abs/1405.2612} {\path{arXiv:1405.2612}},
  \href {http://dx.doi.org/10.1103/PhysRevLett.114.092002}
  {\path{doi:10.1103/PhysRevLett.114.092002}}.

\bibitem{Chien:2015vja}
Y.-T. Chien, A.~Emerman, Z.-B. Kang, G.~Ovanesyan, I.~Vitev, {Jet Quenching
  from QCD Evolution}, Phys. Rev. D93~(7) (2016) 074030.
\newblock \href {http://arxiv.org/abs/1509.02936} {\path{arXiv:1509.02936}},
  \href {http://dx.doi.org/10.1103/PhysRevD.93.074030}
  {\path{doi:10.1103/PhysRevD.93.074030}}.

\bibitem{Baty}
A.~Baty, {Charged particle nuclear modification factor in PbPb at 5.02 TeV with
  CMS}, these proceedings.

\bibitem{Kang:2016ofv}
Z.-B. Kang, F.~Ringer, I.~Vitev, {Effective field theory approach to open heavy
  flavor production in heavy-ion collisions}\href
  {http://arxiv.org/abs/1610.02043} {\path{arXiv:1610.02043}}.

\bibitem{Leibovich:2003jd}
A.~K. Leibovich, Z.~Ligeti, M.~B. Wise, {Comment on quark masses in SCET},
  Phys. Lett. B564 (2003) 231--234.
\newblock \href {http://arxiv.org/abs/hep-ph/0303099}
  {\path{arXiv:hep-ph/0303099}}, \href
  {http://dx.doi.org/10.1016/S0370-2693(03)00565-3}
  {\path{doi:10.1016/S0370-2693(03)00565-3}}.

\bibitem{Jager:2002xm}
B.~Jager, A.~Schafer, M.~Stratmann, W.~Vogelsang, {Next-to-leading order QCD
  corrections to high p(T) pion production in longitudinally polarized pp
  collisions}, Phys. Rev. D67 (2003) 054005.
\newblock \href {http://arxiv.org/abs/hep-ph/0211007}
  {\path{arXiv:hep-ph/0211007}}, \href
  {http://dx.doi.org/10.1103/PhysRevD.67.054005}
  {\path{doi:10.1103/PhysRevD.67.054005}}.

\bibitem{Kniehl:2008zza}
B.~A. Kniehl, G.~Kramer, I.~Schienbein, H.~Spiesberger, {Finite-mass effects on
  inclusive $B$ meson hadroproduction}, Phys. Rev. D77 (2008) 014011.
\newblock \href {http://arxiv.org/abs/0705.4392} {\path{arXiv:0705.4392}},
  \href {http://dx.doi.org/10.1103/PhysRevD.77.014011}
  {\path{doi:10.1103/PhysRevD.77.014011}}.

\bibitem{Chien:2015hda}
Y.-T. Chien, I.~Vitev, {Towards the understanding of jet shapes and cross
  sections in heavy ion collisions using soft-collinear effective theory}, JHEP
  05 (2016) 023.
\newblock \href {http://arxiv.org/abs/1509.07257} {\path{arXiv:1509.07257}},
  \href {http://dx.doi.org/10.1007/JHEP05(2016)023}
  {\path{doi:10.1007/JHEP05(2016)023}}.

\bibitem{Aad:2014bxa}
G.~Aad, et~al., {Measurements of the Nuclear Modification Factor for Jets in
  Pb+Pb Collisions at $\sqrt{s_{\mathrm{NN}}}=2.76$ TeV with the ATLAS
  Detector}, Phys. Rev. Lett. 114~(7) (2015) 072302.
\newblock \href {http://arxiv.org/abs/1411.2357} {\path{arXiv:1411.2357}},
  \href {http://dx.doi.org/10.1103/PhysRevLett.114.072302}
  {\path{doi:10.1103/PhysRevLett.114.072302}}.

\bibitem{Chatrchyan:2013kwa}
S.~Chatrchyan, et~al., {Modification of jet shapes in PbPb collisions at $\sqrt
  {s_{NN}} = 2.76$ TeV}, Phys. Lett. B730 (2014) 243--263.
\newblock \href {http://arxiv.org/abs/1310.0878} {\path{arXiv:1310.0878}},
  \href {http://dx.doi.org/10.1016/j.physletb.2014.01.042}
  {\path{doi:10.1016/j.physletb.2014.01.042}}.

\bibitem{Larkoski:2014wba}
A.~J. Larkoski, S.~Marzani, G.~Soyez, J.~Thaler, {Soft Drop}, JHEP 05 (2014)
  146.
\newblock \href {http://arxiv.org/abs/1402.2657} {\path{arXiv:1402.2657}},
  \href {http://dx.doi.org/10.1007/JHEP05(2014)146}
  {\path{doi:10.1007/JHEP05(2014)146}}.

\bibitem{Chien:2016led}
Y.-T. Chien, I.~Vitev, {Probing the hardest branching of jets in heavy ion
  collisions}\href {http://arxiv.org/abs/1608.07283} {\path{arXiv:1608.07283}}.

\bibitem{KVX2017}
Z.~Kang, I.~Vitev, H.~Xing, {Momentum imbalance of vector boson tagged jets},
  in preparation.

\end{thebibliography}







\end{document}